\documentclass[conference]{IEEEtran}
\usepackage{lineno,hyperref}
\usepackage{multirow}
\usepackage{graphicx}
\usepackage{amsmath}
\usepackage{amssymb}

\ifCLASSINFOpdf
 
\begin{document}

\title{Breast mass segmentation based on ultrasonic entropy maps and attention gated U-Net}


\author{\IEEEauthorblockN{Micha\l{} Byra\IEEEauthorrefmark{1}\IEEEauthorrefmark{2}, Piotr Jarosik\IEEEauthorrefmark{3}, Katarzyna Dobruch-Sobczak\IEEEauthorrefmark{2}, Ziemowit Klimonda\IEEEauthorrefmark{2}, \\Hanna Piotrzkowska-Wr\'oblewska\IEEEauthorrefmark{2}, Jerzy Litniewski\IEEEauthorrefmark{2}, Andrzej Nowicki\IEEEauthorrefmark{2} }

\IEEEauthorblockA{\IEEEauthorrefmark{2}Department of Ultrasound, Institute of Fundamental Technological Research, \\Polish Academy of Sciences, Warsaw, Poland}
\IEEEauthorblockA{\IEEEauthorrefmark{3}Department of Information and Computational Science, Institute of Fundamental Technological Research, \\Polish Academy of Sciences, Warsaw, Poland}
\IEEEauthorblockA{\IEEEauthorrefmark{1}Corresponding author, e-mail: mbyra@ippt.pan.pl}
}

\maketitle

\begin{abstract}
 
We propose a novel deep learning based approach to breast mass segmentation in ultrasound (US) imaging. In comparison to commonly applied segmentation methods, which use US images, our approach is based on quantitative entropy parametric maps. To segment the breast masses we utilized an attention gated U-Net convolutional neural network. US images and entropy maps were generated based on raw US signals collected from 269 breast masses. The segmentation networks were developed separately using US image and entropy maps, and evaluated on a test set of 81 breast masses. The attention U-Net trained based on entropy maps achieved average Dice score of 0.60 (median 0.71), while for the model trained using US images we obtained average Dice score of 0.53 (median 0.59). Our  work presents the  feasibility  of  using  quantitative US parametric  maps  for the breast  mass  segmentation. The obtained results suggest that US parametric maps, which provide the information about local tissue scattering properties, might  be  more suitable for the  development  of breast mass segmentation  methods than regular US images. 

\end{abstract}


\IEEEpeerreviewmaketitle

\section{Introduction}

Breast cancer is the most common invasive cancers in women worldwide \cite{bray2018global}. Ultrasound (US) scanning has been widely used for breast mass diagnosis in clinics. Breast mass segmentation is an important part of breast computer aided diagnosis (CAD) systems. Automatic and accurate mass segmentation enables extraction of handcrafted features for the differentiation of malignant and benign breast masses \cite{wu2019artificial}. However, breast mass segmentation in US images is considered difficult due to speckle noise and blurred breast mass boundaries. Recently, deep learning methods based on convolutional neural networks (CNNs) have been proposed for breast mass detection and segmentation \cite{yap2017automated,10.1117/1.JMI.6.1.011007,doi:10.1002/mp.13268}. These data driven machine learning methods can automatically process US images to determine the segmentation mask.

In US imaging, the appearance of tissues is related to the applied US image reconstruction method. Quantitative US techniques use raw US data (before US image reconstruction) to generate parametric maps illustrating local physical properties of tissues \cite{oelze2016review}. Parametric maps can provide information about tissues that is not present in regular US images \cite{oelze2016review}. In this work, we investigate the feasibility of developing deep learning segmentation methods based on entropy parametric maps, which visualize the information associated with tissue micro-structure \cite{tsui2017small}. Entropy imaging has been successfully used for the breast mass characterization \cite{tsui2017small}. Since the parametric maps are quantitative, we hypothesize that the segmentation based on the parametric maps may provide better performance than in the case of the US images. To segment the breast masses we develop U-Net CNNs equipped with attention gates.  

\section{Materials and Methods}

\subsection{Dataset}

To develop and evaluate deep learning models we used raw US data (before US image reconstruction) collected from 269 breast masses. 123 masses were malignant and 146 masses were benign. Malignant masses were histologically assessed by core needle biopsy, benign masses were assessed either by the biopsy or a two year observation. This retrospective study was approved by the Institutional Review Board. The data were acquired during routing scanning performed at the Maria Sklodowska-Curie Memorial Cancer Centre and Institute of Oncology in Warsaw. Raw US data were collected using the Ultrasonix SonixTouch Research US scanner equipped with the L14-5/38 transducer operating at center imaging frequency of 9 MHz. For each mass two perpendicular scans were performed. Each signal frame consisted of 256 scan lines sampled at 40 MHz. US images were reconstructed based on the raw data. First, Hilbert transform was applied to calculate US signals' amplitudes based on raw US signals. Second, amplitude samples were logarithmically compressed and mapped to 8 bits at a dynamic range of 50 dB. Next, the reconstructed US images were used by an experienced radiologist to outline regions of interest (ROIs) indicating breast mass areas. More information about the imaging protocol can be found in our previous papers \cite{byra2016classification,piotrzkowska2017open}. 

\subsection{Entropy imaging}

Entropy parametric maps were generated using amplitude samples calculated with the Hilbert transform based on raw US signals. To generate the maps we used the sliding window technique proposed by Tsui et al. \cite{tsui2017small}. Square window of size equal to 2 wavelengths (100x14 pixels) was used to collect local amplitude samples to estimate the probability density function and calculate the entropy value using the following equation:

\begin{equation}
    E = - \int f(A) \text{ln} f(A) dA,
\end{equation}

\noindent where $A$ refers to the signal amplitude and  $f(A)$ is the amplitude probability density function. US images and entropy maps generated for two breast masses are presented in Fig. 1. 

\begin{figure}[]
	\begin{center}
		\includegraphics[width=1\linewidth]{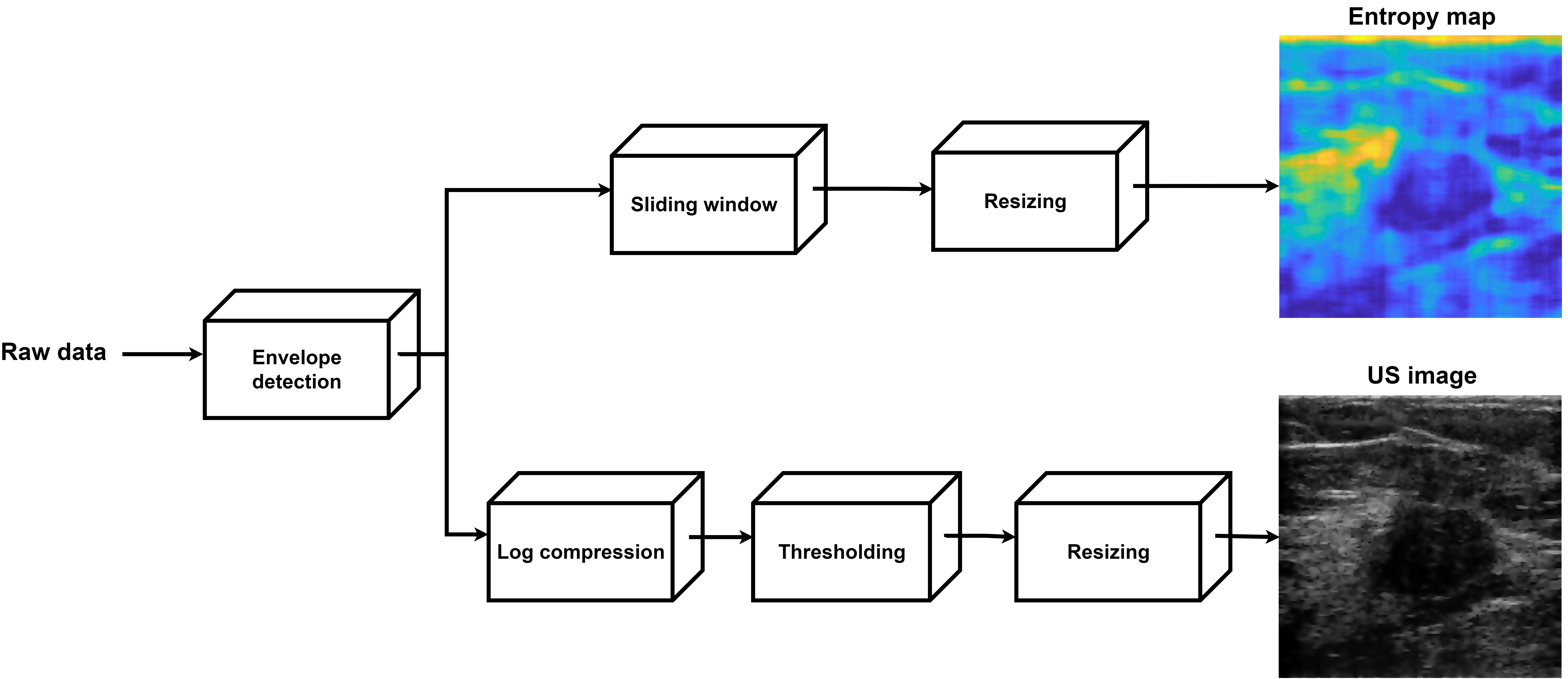}
	\end{center}
	\caption{Pipeline describing the generation of ultrasound images and entropy maps based on raw data collected from breast masses.}
	\label{fig_recon}
\end{figure}

\subsection{Segmentation method}

We used attention gated U-Net CNN to segment breast masses. The scheme of our method is shown in Fig. 2. In comparison to the standard U-Net architecture, the attention U-Net includes attention gates that process feature maps propagated through the skip connections. The aim of the attention gates is to filter the feature maps to improve network's capabilities to focus on important regions in the image \cite{schlemper2019attention,byra2019knee}. Additionally, to improve the performance we applied the following transfer learning technique. The first two convolutional blocks of the attention U-Net CNN were initialized with the weights of the VGG19 network pre-trained on the ImageNet dataset \cite{deng2009imagenet,simonyan2014very}. The VGG19 was originally developed to classify objects from the ImageNet dataset. The first convolutional blocks in a CNN are commonly responsible for the recognition of edges and blobs, therefore thanks to the applied transfer learning technique our U-Net CNN from the very beginning had the ability to recognize low level image features related to local image patterns. To enable transfer learning, US images and entropy maps were resized to the default VGG19 input size of 224x224. The VGG19 CNN was developed for RGB images, but the US images and entropy maps are gray scale. To convert those images to RGB we used the matching layer method \cite{byra2019breast}. The aim of this layer was to rescale pixel intensities and convert gray scale images to RGB color space.

\begin{figure}[]
	\begin{center}
		\includegraphics[width=1\linewidth]{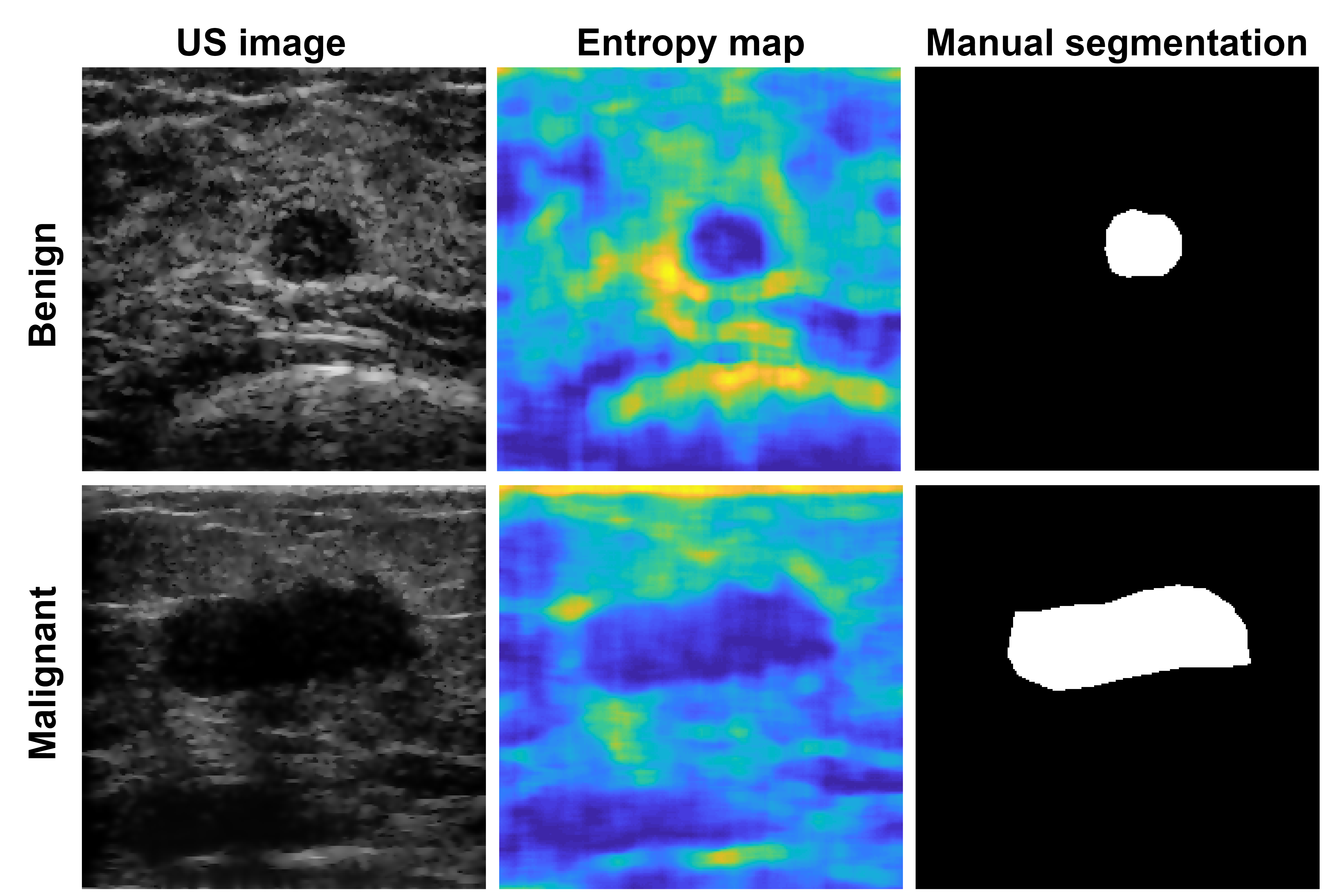}
	\end{center}
	\caption{Ultrasound (US) images and entropy maps generated for sample malignant and benign breast masses, and the corresponding manual segmentations determined based on US images by the radiologist. Entropy parametric maps generated based on raw US signals are commonly less noisy than US images. 
	}
	\label{fig_example}
\end{figure}

\begin{figure*}[t!]
	\centering
	\includegraphics[width=1.0\linewidth]{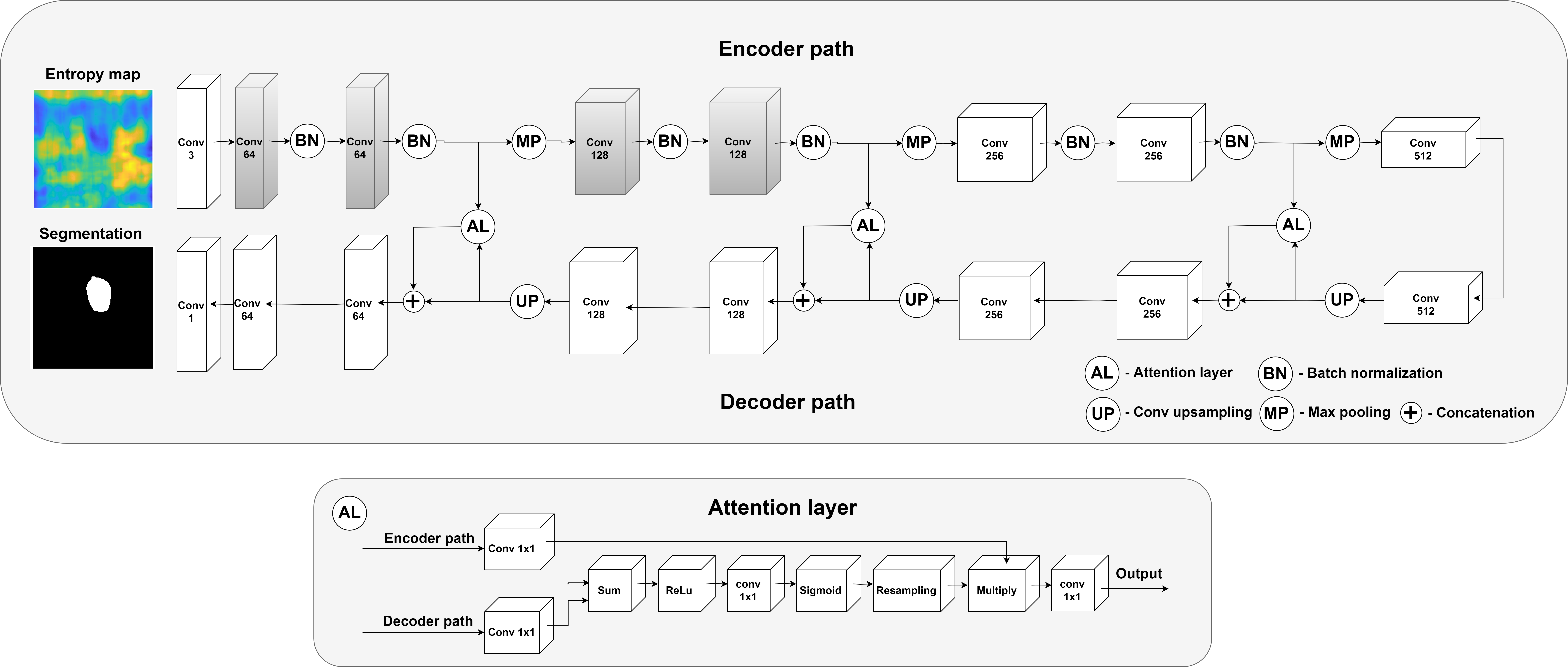}%
	\caption{The architecture of the attention U-Net CNN used for the breast mass segmentation. Convolutional blocks extracted from the VGG19 network are indicated with gray color. AL -– attention layer, BN -- batch normalization, Conv -- 2D convolutional block, MP -- max pooling operation, Up -- up sampling with a 2D transposed convolutional block (kernel size of 2x2, stride of 2x2). Each convolutional block, except for the first and the last block, utilized the rectifier linear unit (ReLu) activation function and 3x3 convolutional filters. The first convolutional block utilized 1D 1x1 convolutional filters (matching layer) without the activation function. The sigmoid activation function was applied for the last convolutional block. }
	\label{fig_net}
	\hfil
\end{figure*}

\subsection{Training and evaluation}

We developed two separate deep learning models. The first CNN was trained using US images, while to train the second one we used entropy maps. The dataset of 269 breast masses was randomly divided into training, validation and test sets with a 147, 41, 81 split (55\%, 15\%, 30\%). The ratio of malignant and benign breast masses was approximately the same for each set. The test set contained data from 38 malignant and 43 benign breast masses. We applied augmentation to improve the training, all US images and entropy maps were horizontally flipped. 

The attention U-Net CNNs were trained to maximize the Dice score based cost function, with the radiologist's ROIs as the ground truth. The Dice coefficient is commonly used for the assessment of segmentation performance, therefore the maximization of this score is desirable. Moreover, the Dice score based cost function is a good choice for imbalanced data, i.e. when the objects, to be segmented, like the breast masses in our case, vary in size \cite{sudre2017generalised}. 

During the training, we monitored the Dice score on the validation set. Each attention U-Net was trained using back-propagation algorithm with the Adam optimizer. The batch size was equal to 16. The learning rate and the momentum were set to 0.0005 and 0.9, respectively. The learning rate was exponentially decreased every 4 epochs by using a drop factor of 0.5 if no improvement was observed on the validation set. The training was stopped after 20 epochs if no improvement in Dice score based cost function was observed on the validation set. After the training was stopped, we selected the better performing models with respect to the validation set for the evaluation. The automatic segmentations calculated for the test set were additionally processed using morphological closing operation with a disk of radius equal to 3 pixels.

\begin{figure*}[t]
	\centering
	\includegraphics[width=6in]{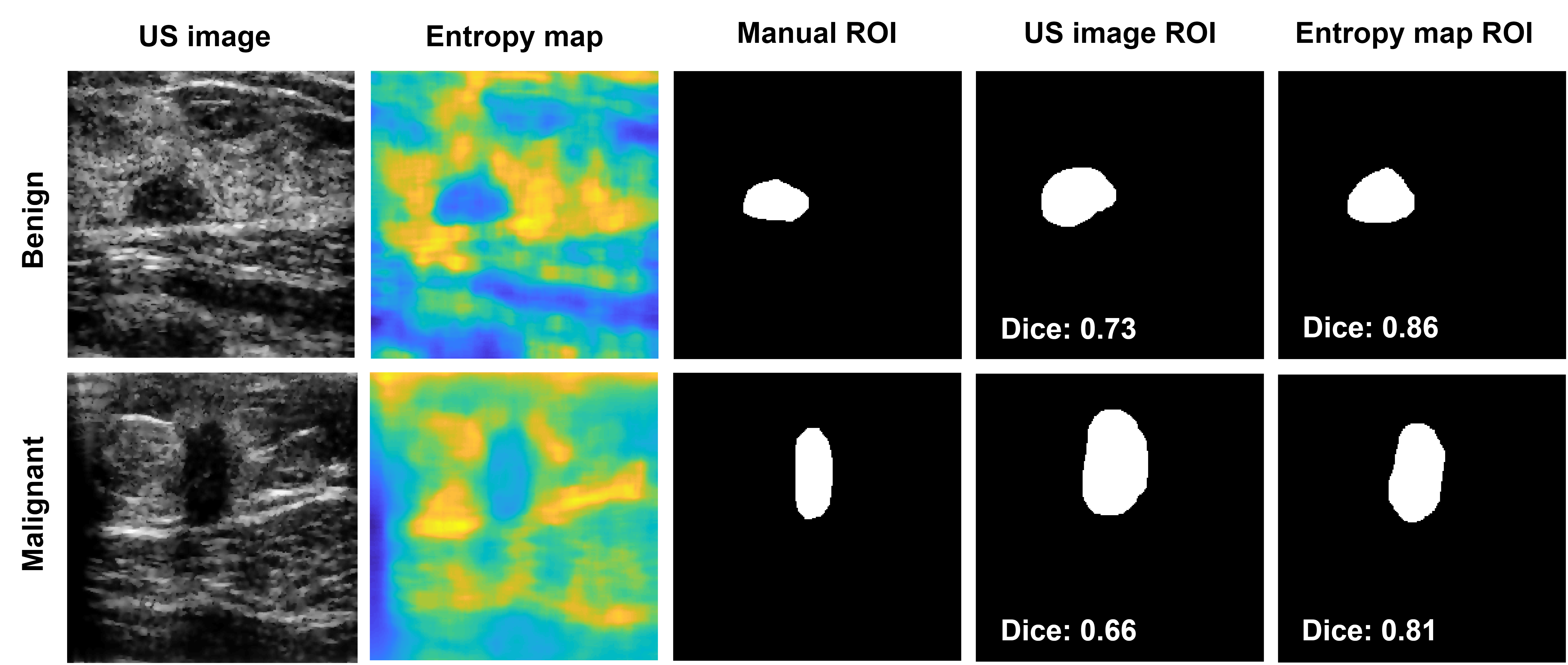}%
	\caption{Segmentation results obtained for the attention U-Net convolutional neural networks developed using ultrasound (US) images and entropy parametric maps. ROI - region of interest.  }
	\label{fig_seg}
	\hfil
\end{figure*}

To evaluate the segmentation performance we calculated the Dice and Jaccard scores using the test set. Wilcoxon rank sum test at significance level of 0.05 was applied to determine whether there were statistically significant differences between the obtained Dice scores. Additionally, we investigated whether there was a difference in performance between the segmentation of malignant and benign breast masses. All calculations were done in Matlab (MathWorks, Inc, USA) and in Python. The networks were implemented in Keras with the Tensorflow backend \cite{abadi2016tensorflow}. The experiments were performed on a computer equipped with a NVIDIA Titan RTX graphics card.

\section{Results}

\begin{table*}[]
    \begin{center}
        \caption{Breast mass segmentation performance achieved by the attention U-Net convolutional neural networks developed using ultrasound (US) images and entropy parametric maps. The average Dice and Jaccard scores (plus median and standard deviation) were calculated based on the test set of 38 malignant and 43 benign breast masses.}
        \begin{tabular}{|c|c|c|c|c|c|c|} 
            \hline 
            Score & Method & Benign & Malignant & Benign and malignant \\
            \hline \hline
            \multirow{2}{*}{Dice} 
            & US image & 0.51 (0.58$\pm$0.29) & 0.54 (0.60$\pm$0.24) & 0.53 (0.59$\pm$0.27) \\  
            & Entropy & 0.60 (0.72$\pm$0.31) & 0.61 (0.70$\pm$0.26) & 0.60 (0.71$\pm$0.29) \\ 
            \hline 
            \multirow{2}{*}{Jaccard} 
            & US image & 0.39 (0.41$\pm$0.25) & 0.41 (0.43$\pm$0.21) & 0.39 (0.42$\pm$0.23) \\ 
            & Entropy & 0.49 (0.56$\pm$0.28) & 0.48 (0.54$\pm$0.25) & 0.49 (0.54$\pm$0.27) \\ 
            \hline
        \end{tabular}
    \end{center}

	\label{tab:results}
\end{table*}

Breast mass segmentation performance achieved by the deep learning models is summarized in Table 1. The attention U-Net developed using entropy maps achieved significantly higher median Dice test score (0.71) than for the US images (0.59). There were no associated differences in segmentation performance between the benign and malignant breast masses, both deep learning models achieved similar scores in this case. Fig. 3 shows  sample automatic segmentations obtained for benign and malignant masses. Boundaries of masses in parametric maps were more visible resulting in better automatic segmentations calculated by the model.

\section{Discussion}

The proposed segmentation method based on the attention U-Net CNN achieved good performance. Our work, for the first time, shows the feasibility of using statistical US parametric maps for the breast mass segmentation. Our results suggest that the parametric maps might be more suitable for the development of segmentation networks than regular US images. For instance, boundaries of breast masses may be more visible in the parametric maps than in US images, making the development of the segmentation method easier. In the case of our study, the attention U-Net trained using entropy maps achieved significantly higher average Dice score, 0.60, than the network developed based on US images, 0.53. 

There are several issues related to our study. First, the ROIs used to train the networks were prepared based on US images. It would be interesting to ask the radiologist to outline the ROIs using entropy maps, and compare the results with the US image based ROIs. Second, the better performance of the network developed using entropy maps could be due to the applied transfer learning technique, which favoured the entropy maps. Third, a research US scanner is required to collect the raw US data necessary for the generation of US parametric maps, which limits the applicability of the proposed method. 

In the future we would like to investigate the usefulness of other US parametric maps for the breast mass segmentation. For instance, it would be interesting to generate attenuation coefficient or Nakagami parameter maps, and use those to train the segmentation CNNs. Moreover, we plan to develop a deep learning model that jointly process different parametric maps to output segmentation ROIs. It would be also interesting to use raw US data directly to develop the segmentation method. 

\section*{Acknowledgement} 

The authors acknowledge grant support from the National Science
Center, Poland (2014/13/B/ST7/01271, 2016/23/B/ST8/03391).

\section*{Conflict of interest statement}

The authors do not have any conflicts of interest. 

\bibliographystyle{IEEEtran}
\bibliography{mybibfile}

\end{document}